\documentclass[12pt]{article}
\usepackage{amssymb}

\def\hybrid{\topmargin 0pt      \oddsidemargin 0pt
        \headheight 0pt \headsep 0pt
        \voffset=-0.5cm
        \hoffset=-0.25in
        \textwidth 6.75in
        \textheight 9.5in       
        \marginparwidth 0.0in
        \parskip 5pt plus 1pt   \jot = 1.5ex}
\catcode`\@=11
\def\marginnote#1{}

\newcount\hour
\newcount\minute
\newtoks\amorpm
\hour=\time\divide\hour by60 \minute=\time{\multiply\hour by60
\global\advance\minute by-\hour}
\edef\standardtime{{\ifnum\hour<12 \global\amorpm={am}%
        \else\global\amorpm={pm}\advance\hour by-12 \fi
        \ifnum\hour=0 \hour=12 \fi
        \number\hour:\ifnum\minute<10 0\fi\number\minute\the\amorpm}}
\edef\militarytime{\number\hour:\ifnum\minute<10 0\fi\number\minute}

\def\draftlabel#1{{\@bsphack\if@filesw {\let\thepage\relax
   \xdef\@gtempa{\write\@auxout{\string
      \newlabel{#1}{{\@currentlabel}{\thepage}}}}}\@gtempa
   \if@nobreak \ifvmode\nobreak\fi\fi\fi\@esphack}
        \gdef\@eqnlabel{#1}}
\def\@eqnlabel{}
\def\@vacuum{}
\def\draftmarginnote#1{\marginpar{\raggedright\scriptsize\tt#1}}
\def\draftlabel#1{{\@bsphack\if@filesw {\let\thepage\relax
   \xdef\@gtempa{\write\@auxout{\string
      \newlabel{#1}{{\@currentlabel}{\thepage}}}}}\@gtempa
   \if@nobreak \ifvmode\nobreak\fi\fi\fi\@esphack}
        \gdef\@eqnlabel{#1}}
\def\@eqnlabel{}
\def\@vacuum{}
\def\draftmarginnote#1{\marginpar{\raggedright\scriptsize\tt#1}}

\def\draft{\oddsidemargin -.5truein
        \def\@oddfoot{\sl preliminary draft \hfil
        \rm\thepage\hfil\sl\today\quad\militarytime}
        \let\@evenfoot\@oddfoot \overfullrule 3pt
        \let\label=\draftlabel
        \let\marginnote=\draftmarginnote
   \def\@eqnnum{(\theequation)\rlap{\kern\marginparsep\tt\@eqnlabel}%
\global\let\@eqnlabel\@vacuum}  }

\def\numberbysection{\@addtoreset{equation}{section}
        \def\theequation{\thesection.\arabic{equation}}}

\def\underline#1{\relax\ifmmode\@@underline#1\else
        $\@@underline{\hbox{#1}}$\relax\fi}

\def\titlepage{\@restonecolfalse\if@twocolumn\@restonecoltrue\onecolumn
     \else \newpage \fi \thispagestyle{empty}\c@page\z@
        \def\thefootnote{\fnsymbol{footnote}} }

\def\endtitlepage{\if@restonecol\twocolumn \else  \fi
        \def\thefootnote{\arabic{footnote}}
        \setcounter{footnote}{0}}  

\numberbysection \hybrid

\newcounter{mo}

\newcommand{\tr}{{\rm tr}}
\newcommand{\ti}[1]{\tilde{#1}}

\newcommand{\mL}{{\mathcal L}}
\newcommand{\mM}{{\mathcal M}}
\newcommand{\mF}{{\mathcal F}}
\newcommand{\mH}{{\mathcal H}}

\newcommand{\vf}{\varphi}
\newcommand{\al}{\alpha}
\newcommand{\be}{\beta}
\newcommand{\ga}{\gamma}
\newcommand{\om}{\omega}
\newcommand{\vth}{\vartheta}

\newcommand{\Mat}{ {\rm Mat}(N,\mathbb C) }

\newcommand{\mC}{\mathbb C}
\newcommand{\mZ}{\mathbb Z}

\newcommand{\ox}{\otimes}

\def\beq{\begin{equation}}
\def\eq{\end{equation}}
\def\p{\partial}

\begin{document}

\setcounter{page}{1}

\date{}
\date{}


\

\begin{center}
%
{\Large{R-matrix-valued Lax pairs and long-range spin chains}}

\vspace{18mm}

{\large  {I. Sechin}\,\footnote{Steklov Mathematical Institute of
Russian Academy of Sciences, 8 Gubkina St., Moscow 119991, Russia;
e-mail: shnbuz@gmail.com}
 \quad\quad\quad\quad
{A. Zotov}\,\footnote{Steklov Mathematical Institute of Russian
Academy of Sciences, 8 Gubkina St., Moscow 119991, Russia;
  e-mail:
zotov@mi.ras.ru}
 }
\end{center}

\vspace{1mm}


 \begin{abstract}
In this paper we discuss $R$-matrix-valued Lax pairs for ${\rm
sl}_N$ Calogero-Moser model and their relation to integrable quantum
long-range spin chains of the Haldane-Shastry-Inozemtsev type.
First, we construct the $R$-matrix-valued Lax pairs for the third
flow of the classical Calogero-Moser model. Then we notice that the
scalar parts (in the auxiliary space) of the $M$-matrices
corresponding to the second and third flows have form of special
spin exchange operators. The freezing trick restricts them to
quantum Hamiltonians of long-range spin chains. We show that for a
special choice of the $R$-matrix these Hamiltonians reproduce those
for the Inozemtsev chain. In the general case related to the
Baxter's elliptic $R$-matrix we obtain a natural anisotropic
extension of the Inozemtsev chain. Commutativity of the Hamiltonians
is verified numerically. Trigonometric limits lead to the
Haldane-Shastry chains and their anisotropic generalizations.
 \end{abstract}



\vspace{10mm}

\small{


 \setcounter{section}{1}

 \setcounter{equation}{0}
\paragraph{Introduction.} Integrable systems
 are known to be actively engaged in high energy physics.
For example, the low energy sector of SUSY (${\mathcal N}=2$) gauge
theories is described by the Seiberg-Witten \cite{SW} solution in
terms of the classical integrable models \cite{GKMMM}, while their
quantum counterparts are described by the supersymmetric vacua of
this gauge theory (deformed by the $\Omega$-background) \cite{NS}. A
link to the conformal field theories is given by the AGT relation
\cite{AGT}, which (in the Nekrasov-Shatashvili limit) turns into a
certain interrelation between integrable systems known as the
spectral duality \cite{MMZZ}. On the CFT side integrable systems
appear also naturally from the Matsuo-Cherednik construction for the
Knizhnik-Zamolodchikov equations \cite{KZ}. Its classical version --
the quantum-classical duality  -- provides a link between the
quantum spin chains and classical many-body integrable systems
\cite{GZZ}. In this paper we discuss alternate example of a relation
between quantum spin chains and classical integrable systems, which
is based on the so-called $R$-matrix-valued Lax pairs
\cite{Inoz0,HW2}, \cite{LOZ,GZ}.

The completely integrable Hamiltonian models can be subdivided into
two large families. The first one consists of many-body systems
including their spin and/or multispin generalizations. A
representative example is given by the Calogero-Moser model. The
classical spinless $N$-body elliptic ${\rm gl}_N$ model is described
by the Hamiltonian
  \beq\label{q02}
  \begin{array}{c}
  \displaystyle{
H_2=\sum\limits_{i=1}^N\frac{p_i^2}{2}-\nu^2\sum\limits_{i<j}^N\wp(q_i-q_j)\,,
 }
 \end{array}
 \eq
where $\wp(x)$ is the Weierstrass $\wp$-function and $\nu\in\mC$ is
a
 coupling constant.
Its spin quantum analogue \cite{GH} is given by
  \beq\label{qq021}
  \begin{array}{c}
  \displaystyle{
 {\hat H}=\sum\limits_{i=1}^N\frac{\hat{p}_i^2}{2}-\sum\limits_{i<j}^N\nu(\nu+\hbar P_{ij})\wp(q_i-q_j)\,,
 }
 \end{array}
 \eq
where $\hat{p}_i=\hbar\p_{q_i}$ and $P_{ij}$ is the (spin exchange)
permutation operator.

The second family of integrable models is represented by integrable
tops, spin chains and/or Gaudin models. In contrast to the previous
family these are governed by numerical (non-dynamical) $R$-matrices.
A typical example is given by the (local) XYZ spin chain Hamiltonian
\cite{Baxter,FT,Skl}
  \beq\label{qq022}
  \begin{array}{c}
  \displaystyle{
 {\hat
 H}=\sum\limits_{i=1}^{N-1}\hat{h}_{i,i+1}+\hat{h}_{N,1}\,,\quad\
 \hat{h}_{i,i+1}=\sum\limits_{a=0}^3\stackrel{i}{\sigma}_a\stackrel{i+1}{\sigma}_a
 J_a\,,
 }
 \end{array}
 \eq
where $J_a$ are constants (anisotropy parameters) and
$\stackrel{i}{\sigma}_a$ are the $a$-th components of the spin
operator at $i$-th site. For the $1/2$-spin case these are the Pauli
matrices acting on the $j$-th tensor component of ${\mathcal
H}=(\mC^2)^{\otimes N}$ -- the Hilbert space of the model. When all
the constants are equal to each other $J_a=J_b$ we have
$\hat{h}_{i,i+1}=2P_{i,i+1}$. Then (\ref{qq022}) is the isotropic
(XXX) Heisenberg Hamiltonian.

In this letter we deal with the integrable models which can be
regarded as an intermediate link between the above mentioned
families. These are the Haldane-Shastry-Inozemtsev long-range spin
chains \cite{HS,Inoz0}. The Hamiltonian
  \beq\label{qq023}
  \begin{array}{c}
  \displaystyle{
 {\hat
 H}=\sum\limits_{i<j}^{N}P_{ij}\,\wp(x_i-x_j)\,,\quad x_i=i/N\,,\ i=1,...,N
 }
 \end{array}
 \eq
describes pairwise interaction of $N$ spins on a unit circle with
equidistant positions. It can be shown that the scaling limit of
(\ref{qq023}) provides the XXX Heisenberg model with the
nearest-neighbor interaction likewise the (periodic) Toda chain is
obtained from the Calogero-Moser model (\ref{q02})
\cite{Inoz-Toda}\footnote{In fact, the Toda model can be also
treated as an intermediate link between the two families since it
admits two types of the Lax representations: $2\times 2$ as for the
spin chains and $N\times N$ as for the many-body systems
\cite{FT}.}. On the other hand the Hamiltonian (\ref{qq023}) can be
obtained from the quantum spin Calogero-Moser one (\ref{qq021}) by
the so-called freezing trick \cite{Polych1}, when the particles
positions are frozen as $q_i\rightarrow x_i$. The models of the
Haldane-Shastry-Inozemtsev type found applications in the AdS/CFT
correspondence, where the problem of computation of anomalous
dimensions of certain ${\mathcal N}=4$ composite operators
emerged.
 The one-loop anomalous dimensions of these operators were calculated by means of the Bethe ansatz method
 for the Heisenberg chain \cite{MZ}, and the higher-loop dilatation operator appeared to be expressed through the
 conserved charges of the long-range chains \cite{ads}.

{\em The purpose of the paper} is to show that the
Haldane-Shastry-Inozemtsev spin chains admit aniso\-tro\-pic
integrable extensions much as XYZ model generalizes the XXX
Heisenberg chain. That is to say that we are going to define an
integrable model with the Hamiltonian of the form
  \beq\label{qq024}
  \begin{array}{c}
  \displaystyle{
 {\hat
 H}=\sum\limits_{i<j}^{N}\sum\limits_{a=0}^3\stackrel{i}{\sigma}_a\stackrel{j}{\sigma}_a
 J_a(x_i-x_j)\,.
 }
 \end{array}
 \eq
To construct such Hamiltonian we use the $R$-matrix-valued Lax pair
\cite{LOZ} for the (spinless) Calogero-Moser model (\ref{q02}). It
is a generalization of the well-known Lax pair with spectral
parameter \cite{Krich1} to the case when the matrix elements are not
scalar functions but $R$-matrices satisfying associative Yang-Baxter
equation \cite{FK,Pol} and some additional properties. The Lax
equations
  \beq\label{qq025}
  \begin{array}{c}
  \displaystyle{
 \mL=[\mL,\mM]
 }
 \end{array}
 \eq
with the Lax pair (\ref{q20})-(\ref{q22}) are equivalent to the
classical equations of motion for the model (\ref{q02}). The matrix
elements of $\mL$, $\mM$ are operators on the Hilbert space
$\mathcal H$, i.e. $\mL,\mM\in\Mat\otimes{\rm End}(\mathcal H)$. We
will refer to $\Mat$ component as the auxiliary space, and to the
${\rm End}(\mathcal H)$ component as the quantum (spin chain) space.

Our strategy is as follows.
At the level of the classical Calogero-Moser model (\ref{q02}) the
above mentioned freezing trick turns into the set of conditions
  \beq\label{qq026}
  \begin{array}{c}
  \displaystyle{
 p_i=0\,,\quad q_i=x_i\,,
 }
 \end{array}
 \eq
understood to be the equilibrium position. Being restricted to the
constraints (\ref{qq026}) the Lax equations (\ref{qq025}) become
$[\mL,\mM]=0$ on-shell (\ref{qq026}). At the same time the
$\mM$-matrix (\ref{q21}) contains the part $\Delta\mM=1_{N\times
N}\otimes \nu\mF^0$ (\ref{q22}), which is a scalar operator in the
auxiliary space. Therefore, we may interpret the reduced Lax
equations as follows:
  \beq\label{qq027}
  \begin{array}{c}
  \displaystyle{
 [\nu\mF^0,\mL]=[\mL,\mM-\Delta\mM]\quad \hbox{on-shell}\
 (\ref{qq026})\,,
 }
 \end{array}
 \eq
where the commutator in the l.h.s. is in the quantum space only.
Thus the $\nu\mF^0$ term is the quantum spin chain Hamiltonian. We
will show that it is of the form (\ref{qq024}), and reproduces the
Inozemtsev chain (\ref{qq023}) for a special choice of the
$R$-matrix.

Unfortunately, in the general (elliptic) case the Lax equations
allow to compute the higher Hamiltonians for the classical model
(\ref{q02}) only but not for the quantum spin chain (\ref{qq027}),
because the Hamiltonian in the latter case appeared as a scalar (in
the auxiliary space) part of $\mM$. Nevertheless we may repeat the
above-described computation procedure to the higher flow of the
Calogero-Moser model.
We will construct the $R$-matrix-valued Lax pair for the third flow.
Then restrict it to the equilibrium position (\ref{qq026}) and find
the scalar (in the auxiliary space) part of the corresponding
$\mM$-matrix. At last, we verify by numerical calculations that the
 Hamiltonian obtained in this way indeed commutes with the one related to the second
 flow (\ref{qq027}).

 %
\paragraph{Classical Calogero-Moser model.}
 \setcounter{equation}{0}
 In this paper we deal with the classical
Calogero-Moser-Sutherland models \cite{Ca,OP}. Equations of motion
  \beq\label{q01}
  \begin{array}{c}
  \displaystyle{
 {\dot q}_i=p_i\,,\quad  {\ddot q}_i=\nu^2\sum\limits_{k: k\neq i}^N\wp'(q_{i}-q_k)\,.
 }
 \end{array}
 \eq
are generated by the Hamiltonian (\ref{q02}) (and the canonical
Poisson brackets $\{p_i,q_j\}=\delta_{ij}$). In the trigonometric
limit $\wp(x)\rightarrow {\pi^2}/{\sin^2(\pi x)}$ the classical
Sutherland model is reproduced.

The Hamiltonian (\ref{q02}) is included into a family of the higher
integrals of motion, which are in involution with respect to the
canonical  Poisson brackets: $\{H_k,H_l\}=0$. For example, the third
Hamiltonian
  \beq\label{q03}
  \begin{array}{c}
  \displaystyle{
 H_3=\sum\limits_{i=1}^N\frac{p_i^3}{3}-\nu^2\sum\limits_{i\neq j}^N p_i\,\wp(q_i-q_j)
 }
 \end{array}
 \eq
 provides equations of motion
  \beq\label{q04}
   \left\{
  \begin{array}{l}
  \displaystyle{
 \p_{t_3}{ q}_i=p_i^2-\nu^2\sum\limits_{k\neq i}\wp(q_{i}-q_k)\,,
 }
 \\
  \displaystyle{
 \p_{t_3}{ p}_i=\nu^2\sum\limits_{k\neq i} (p_i+p_k)\wp'(q_{i}-q_k)\,.
 }
 \end{array}
 \right.
 \eq
All the flows are described by the Lax equations
  \beq\label{q05}
  \begin{array}{c}
  \displaystyle{
\p_{t_k}{L}(z)\equiv\{H_k,L(z)\}=[L(z),M^{(k)}(z)]\,,
 }
 \end{array}
 \eq
 where $L(z)$ and $M^{(k)}(z)$ are $N\!\times\! N$ matrices depending on
 the spectral parameter $z$, which does not enter equations
 of motion. So that (\ref{q05}) are identities in $z$ on the equations
 motions.
 The Lax matrix is as follows \cite{Krich1}:
  \beq\label{q06}
  \begin{array}{c}
  \displaystyle{
L_{ij}(z)=\delta_{ij}p_i+\nu(1-\delta_{ij})\,\phi(z,q_{ij})\,,\quad
q_{ij}=q_i-q_j\,,\quad
\phi(z,q)=\frac{\vth'(0)\vth(q+z)}{\vth(q)\,\vth(z)}\,,
 }
 \end{array}
 \eq
 where $\vth(z)$ is the odd Riemann theta-function\footnote{In the trigonometric
 limit  $\phi(z,q)\rightarrow \pi(\cot(\pi z)+\cot(\pi
 q))$.}. The $M^{(2)}$-matrix is of the form
  \beq\label{q07}
  \begin{array}{c}
  \displaystyle{
M^{(2)}_{ij}(z)=\nu d_i\delta_{ij}
+\nu(1-\delta_{ij})f(z,q_{ij})\,,\quad d_i=-\sum\limits_{k\neq i}^N
f(0,q_{ik})\,,
 }
 \end{array}
 \eq
where $f(z,q)=\p_q\phi(z,q)$, and $f(0,q)$ coincides with $-\wp(q)$
up to a constant. Namely,
  \beq\label{q071}
  \begin{array}{c}
  \displaystyle{
f(0,z)=\p_z^2\log\vth(z)=-\wp(z)+\frac{1}{3}\frac{\vth'''(0)}{\vth'(0)}\,.
 }
 \end{array}
 \eq
For the third flow\footnote{While the Hamiltonians $H_k$ are
evaluated from $\tr L^k(z)$, the expressions for $M^{(k)}(z)$
corresponding to higher flows can be similarly extracted from $\tr_2
(r_{12}(z,w)L^{k-1}_2(w))$, where $r_{12}(z,w)$ is the classical
$r$-matrix.} (\ref{q03})-(\ref{q04}) the $M$-matrix is of the form:
  \beq\label{q08}
  \begin{array}{c}
  \displaystyle{
M^{(3)}_{ij}(z)=-\delta_{ij}\,\nu \sum\limits_{k\neq i}(p_i + p_k)
f(0,q_{ik}) +
 }
 \\ \ \\
  \displaystyle{
 +(1-\delta_{ij})\Big( \nu (p_i + p_j) f(z, q_{ij})
 + \nu^2\sum\limits_{k\neq i,j} (\phi(z, q_{ik}) f(z, q_{kj}) - \phi(z, q_{ij})f(0,q_{kj}))
 \Big)\,.
            }
 \end{array}
 \eq
Verification of the above statements is based on the identities
  \beq\label{q09}
  \begin{array}{c}
  \displaystyle{
\phi(z,q_{ab})f(z,q_{ba})-f(z,q_{ab})\phi(z,q_{ba})=\wp'(q_{ab})\,,
 }
 \end{array}
 \eq
  \beq\label{q101}
  \begin{array}{c}
  \displaystyle{
\phi(z,q_{ab})f(z,q_{bc})-f(z,q_{ab})\phi(z,q_{bc})=
\phi(z,q_{ac})(f(0,q_{bc})-f(0,q_{ab}))\,,
 }
 \end{array}
 \eq
which follow from
  \beq\label{q10}
  \begin{array}{c}
  \displaystyle{
\phi(z,q)\phi(z,-q)=\wp(z)-\wp(q)=f(0,q)-f(0,z)
 }
 \end{array}
 \eq
and the (genus one) Fay identity for the Kronecker function
$\phi(z,w)$ \cite{Weil}:
  \beq\label{q11}
  \begin{array}{c}
  \displaystyle{
\phi(z,q_{ab})\phi(w,q_{bc})=\phi(w,q_{ac})\phi(z-w,q_{ab})+\phi(w-z,q_{bc})\phi(z,q_{ac})\,.
 }
 \end{array}
 \eq
\paragraph{Haldane-Shastry-Inozemtsev chain}
 \setcounter{section}{2}
 \setcounter{equation}{0}
 \cite{HS,Inoz0}. The Hamiltonian of the Inozemtsev chain has form
  \beq\label{q12}
  \begin{array}{c}
  \displaystyle{
 H_2^{\mathrm{Inoz}} = \sum_{i < j} P_{ij}\,\wp(x_i-x_j)\,.
 }
 \end{array}
 \eq
In the trigonometric limit it reproduces the Haldane-Shastry model:
  \beq\label{q13}
  \begin{array}{c}
  \displaystyle{
H_2^{\mathrm{HS}} = \sum_{i < j} \frac{P_{ij}}{\sin^2{\pi
(x_i-x_j)}}\,.
 }
 \end{array}
 \eq
In (\ref{q12})-(\ref{q13}) $P_{ij}$ is the permutation
(or spin exchange) operator, which acts on the Hilbert space
$(\mC^2)^{\otimes N}$ of the chain. It interchanges the $i$-th and
$j$-th components in the tensor product (and keeps unchanged the
rest of the components)\footnote{In the general case
 $P_{ij}=\sum\limits_{a,b=1}^{\ti N}\stackrel{i}{E}_{ab}\stackrel{j}{E}_{ba}$, where $\{E_{ab}\in{\rm Mat}_{\ti N},\ a,b=1...{\ti N}\}$ --
is the standard basis in ${\rm Mat}_{\ti N}$:
$(E_{ab})_{cd}=\delta_{ac}\delta_{bd}$. In (\ref{q12})-(\ref{q131})
$\ti N=2$.}:
  \beq\label{q131}
  \begin{array}{c}
  \displaystyle{
 P_{12}=\frac12\sum\limits_{\al=0}^3
 \sigma_\al\otimes\sigma_\al\equiv\frac12
 \sum\limits_{\al=0}^3\stackrel{1}{\sigma}_\al\stackrel{2}{\sigma}_\al\,,
 }
 \end{array}
 \eq
 where $\sigma_\al$ are the Pauli matrices.
The positions $x_j$ are fixed and equidistant:
  \beq\label{q14}
  \begin{array}{c}
  \displaystyle{
x_j=\frac{j}N\,,\quad j=1,...,N\,.
 }
 \end{array}
 \eq
The model admits the quantum Lax representation \cite{Inoz0}:
  \beq\label{q15}
  \begin{array}{c}
  \displaystyle{
[H_2^{\mathrm{Inoz}},\mL^{\mathrm{Inoz}}(z)]=[\mL^{\mathrm{Inoz}}(z),-\mM^{\mathrm{Inoz}}(z)]
 }
 \end{array}
 \eq
 with
  \beq\label{q16}
  \begin{array}{c}
  \displaystyle{
\mL^{\mathrm{Inoz}}(z)=\sum\limits_{i,j}E_{ij}\otimes
(1-\delta_{ij})P_{ij}\,\phi(z,x_{ij})\,,\quad x_{ij}=x_i-x_j
 }
 \end{array}
 \eq
and\footnote{In \cite{Inoz0} $\mM^{\mathrm{Inoz}}$ has different
sign.}
  \beq\label{q17}
  \begin{array}{c}
  \displaystyle{
\mM^{\mathrm{Inoz}}(z)=\sum\limits_{i,j}E_{ij}\otimes\Big(
d_i\delta_{ij} +(1-\delta_{ij})P_{ij}\,f(z,x_{ij})\Big)\,,\quad
d_i=-\sum\limits_{k\neq i}^N P_{ik}f(0,x_{ik})\,.
 }
 \end{array}
 \eq
So that the Lax matrix is of $N\times N$ size with matrix elements
being proportional to the permutation operators $P_{ij}\in{\rm
Mat}(2^N)$. Therefore, $\mL,\mM\in{\rm Mat}(N2^N)$.

The Lax pair (\ref{q16})-(\ref{q17}) owes its origin to the quantum
spin Calogero-Moser model \cite{Polych0,HW}. The long-range spin
chain appears after imposing (\ref{q14}), which is treated as the
freezing trick in the spin Calogero-Moser model \cite{Polych1}. In
classical mechanics (\ref{q14}) means that there is an equilibrium
position, where the particles coordinates are fixed as $q_j=x_j$ and
$p_j=0$.

 In the general case the Lax equation (\ref{q15})
does not allow to calculate the higher integrals of motion. This
becomes possible in special cases when the sum up to zero condition
($\sum_i \mM_{ij}=\sum_j \mM_{ij}=0$) is fulfilled. In the latter
case the higher conserved quantities appear from the total sum of
elements of powers of $\mL$. In our case the Lax pair is elliptic,
and there is no such condition. The receipt for higher integrals was
conjectured in \cite{Inoz0} and then discussed (and partly proved)
in \cite{Inoz1}. Two next Hamiltonians commuting with (\ref{q12})
are of the form:
  \beq\label{q18}
  \begin{array}{c}
  \displaystyle{
J_1 = { \sum\limits_{i,j,k}}'
    \left( E_1(x_{ij}) + E_1(x_{jk}) + E_1(x_{ki}) \right) [P_{ij},
    P_{jk}]\,,
 }
 \end{array}
 \eq
  \beq\label{q19}
  \begin{array}{c}
  \displaystyle{
J_2 = { \sum\limits_{i,j,k}}'\left(
        2 \left( E_1(x_{ij}) + E_1(x_{jk}) + E_1(x_{ki}) \right)^3 +
        \wp'(x_{ij}) + \wp'(x_{jk}) + \wp'(x_{ki})
    \right) [P_{ij}, P_{jk}]\,,
 }
 \end{array}
 \eq
 where a prime means that the corresponding summation is over all
 not coincident values of indices, and $E_1(z)=\p_z\log\vth(z)$.


\paragraph{R-matrix-valued Lax pairs.}
 \setcounter{section}{3}
 \setcounter{equation}{0}
In \cite{LOZ} (see also \cite{GZ}) the following generalization of
the Lax pair (\ref{q06})-(\ref{q07}) for the classical
Calogero-Moser model was suggested\footnote{In fact, a similar Lax
pair was proposed in \cite{HW2} for the quantum trigonometric spin
Calogero-Sutherland model. In that paper the $R$-matrix was chosen
to be the classical trigonometric one (i.e. the corresponding Lax
pair was without spectral parameter as it is for the ordinary Lax
pair of the Sutherland model) for $\ti N=2$ case.}
  \beq\label{q20}
  \begin{array}{c}
    \displaystyle{
{\mL}(z)=\sum\limits_{i,j=1}^N E_{ij}\otimes \mL_{ij}(z)\,,
 \quad\quad
\mL_{ij}(z)= 1_{\ti N}^{\otimes N}\, \delta_{ij}p_i
+\nu(1-\delta_{ij})R^{\,z}_{ij}(q_{ij})
 }
 \end{array}
 \eq
and similarly
  \beq\label{q21}
  \begin{array}{c}
  \displaystyle{
\mM^{(2)}_{ij}(z)=\nu d_i\delta_{ij}
+\nu(1-\delta_{ij})F^{\,z}_{ij}(q_{ij})+\nu\delta_{ij}\,\mF^{\,0}\,,\quad
d_i=-\sum\limits_{k: k\neq i}^N F^{\,0}_{ik}(q_{ik})\,,
 }
 \end{array}
 \eq
  \beq\label{q22}
  \begin{array}{c}
  \displaystyle{
\mF^{\,0}
 =\sum\limits_{k<m}^N F^{\,0}_{km}(q_{km})
 =\frac{1}{2}\sum\limits_{k,m=1}^N F^{\,0}_{km}(q_{km})\,.
 }
 \end{array}
 \eq
where $F^{\,z}_{ij}(q)=\p_q R^{\,z}_{ij}(q)$. By the construction
${\mL}(z),{\mM}^{(2)}(z)\in{\rm Mat}(N {\ti N}^N)$. The Lax equation
${\dot \mL}=[\mL,\mM^{(2)}]$ is equivalent to (\ref{q01}) with the
coupling constant ${\ti N}\nu$ instead of $\nu$. For exact matching
one should rescale $\nu\rightarrow\nu/{\ti N}$ in
(\ref{q20})-(\ref{q22}) but we keep it as it is.

The Lax pair (\ref{q20})-(\ref{q22}) is called $R$-matrix-valued Lax
pair since it can be viewed as $N\times N$ matrices which matrix
elements are quantum ${\rm GL}_{\ti N}$ $R$-matrices (or its
derivatives), satisfying the associative Yang-Baxter equation
\cite{FK}
  \beq\label{q23}
  \begin{array}{c}
  \displaystyle{
 R^z_{ab}
 R^{w}_{bc}=R^{w}_{ac}R_{ab}^{z-w}+R^{w-z}_{bc}R^z_{ac}\,,\
 \ R^z_{ab}=R^z_{ab}(q_a\!-\!q_b)\,.
 }
 \end{array}
 \eq
It was observed in \cite{Pol} that (\ref{q23}) is fulfilled by the
elliptic Baxter-Belavin \cite{Baxter,Belavin} $R$-matrix (written in
proper normalization):
 \beq\label{q24}
 \begin{array}{c}
  \displaystyle{
R_{12}^z(q)=\sum\limits_a T_a\otimes T_{-a}
 \exp\left(2\pi\imath\frac{a_2}{\ti N}\,q\right)\,\phi\left(q,z+\frac{a_1+a_2\tau}{{\ti
N}}\right)\,,\quad
 a=(a_1,a_2)\in \mZ_{\ti N}\times\mZ_{\ti N}\,,
  }
 \end{array}
 \eq
where the basis $T_a$ is defined in terms of the finite dimensional
representation of the Heisenberg group
 \beq\label{q25}
 \begin{array}{c}
  \displaystyle{
 T_a=T_{a_1 a_2}=\exp\left(\frac{\pi\imath}{{\ti N}}\,a_1
 a_2\right)Q^{a_1}\Lambda^{a_2}\,,\quad
 a=(a_1,a_2)\in\mZ_{\ti N}\times\mZ_{\ti N}
 }
 \\
  \displaystyle{
Q_{kl}=\delta_{kl}\exp\left(\frac{2\pi
 \imath}{{\ti N}}k\right)\,,\ \ \ \Lambda_{kl}=\delta_{k-l+1=0\,{\hbox{\tiny{mod}}}
 {\ti N}}\,,\quad Q^{\ti N}=\Lambda^{\ti N}=1_{{\ti N}}\,.
 }
 \end{array}
 \eq
Equations (\ref{q23}) are matrix analogues of the Fay identities
(\ref{q11}) (they coincide for $\ti N=1$). In the same way the
unitarity property\footnote{Different properties and identities of
the Baxter-Belavin $R$-matrix similar to the elliptic function
identities can be found also in \cite{LOZ15,Unitary}.}
  \beq\label{q26}
  \begin{array}{c}
  \displaystyle{
 R^z_{12}(q_{12}) R^{z}_{21}(q_{21})=1_{\ti N}\otimes 1_{\ti N}\,{\ti N}^2(\wp({\ti N}z)-\wp(q_{12}))
 }
 \end{array}
 \eq
is similar to (\ref{q10}). Together with the skew-symmetry (likewise
$\phi(z,q)=-\phi(-z,-q)$)
  \beq\label{q261}
  \begin{array}{c}
  \displaystyle{
 R^z_{12}(q)=-R^{-z}_{21}(-q)
 }
 \end{array}
 \eq
 equations (\ref{q23}) and (\ref{q26}) results to the quantum
Yang-Baxter equation
  \beq\label{q27}
  \begin{array}{c}
  \displaystyle{
 R^z_{ab}(q_{ab})R^z_{ac}(q_{ac})R^z_{bc}(q_{bc})=R^z_{bc}(q_{bc})R^z_{ac}(q_{ac})R^z_{ab}(q_{ab})\,.
 }
 \end{array}
 \eq
Coming back to the Lax pair (\ref{q20})-(\ref{q21}) it must be
emphasized that it is a straightforward generalization of the
Krichever's Lax pair (\ref{q06})-(\ref{q07}) except the last term
(\ref{q22}), which is not necessary in (\ref{q07}) since for $\ti
N=1$ case it is proportional to the identity matrix. The matrix
function $F^0(q)$ entering this term is simply related to the
classical $r$-matrix\footnote{The classical limit is of the form
$R_{12}^z(q)=\frac{1\otimes 1}{z}+r_{12}(q)+O(z)$, and the classical
$r$-matrix is skew-symmetric  $r_{12}(q)=-r_{21}(-q)$.}:
  \beq\label{q28}
  \begin{array}{c}
  \displaystyle{
F^{\,0}_{ij}(q)=F^{\,z}_{ij}(q)|_{z=0}=\p_qr_{ij}(q)=F^{\,0}_{ji}(-q)\,.
 }
 \end{array}
 \eq
From the above it follows that we also have $R$-matrix analogues for
identities (\ref{q09}), (\ref{q101}):
  \beq\label{q29}
  \begin{array}{c}
  \displaystyle{
 R^z_{ab} F^z_{ba} - F^z_{ab} R^z_{ba}={\ti N}^2\wp'(q_{ab})\,,
 }
 \end{array}
 \eq
  \beq\label{q30}
  \begin{array}{c}
  \displaystyle{
R^z_{ab} F^z_{bc} - F^z_{ab} R^z_{bc} = F^0_{bc} R^z_{ac} - R^z_{ac}
F^0_{ab}\,.
 }
 \end{array}
 \eq
The latter identities underly the Lax equations for
(\ref{q20})-(\ref{q22}). The role of the $\mF^0$ term is to correct
the order of multipliers
  \beq\label{q31}
  \begin{array}{c}
  \displaystyle{
[R^z_{ac},\mF^{\,0}]+\sum\limits_{b\neq a,c}R^z_{ab} F^z_{bc} -
F^z_{ab} R^z_{bc} = \sum\limits_{b\neq c} R^z_{ac} F^0_{bc} -
 \sum\limits_{b\neq a} F^0_{ab} R^z_{ac}\,.
 }
 \end{array}
 \eq
%
It is natural to expect existence of higher $R$-matrix-valued
$M$-matrices related to higher Hamiltonians (\ref{q05}). Here we
propose $R$-matrix-valued generalization of the $M$-matrix for the
third flow (\ref{q08}). It is of the form:
  \beq\label{q32}
  \begin{array}{c}
  \displaystyle{
\mM^{(3)}_{ij}(z)=-\delta_{ij}\,\nu \sum\limits_{k\neq i}(p_i + p_k)
F^0_{ik}(q_{ik}) +
 }
 \\ \ \\
  \displaystyle{
 +(1-\delta_{ij})\Big( \nu (p_i + p_j) F^z_{ij}(q_{ij})
 + \nu^2\sum\limits_{k\neq i,j} (R_{ik}^z(q_{ik}) F^z_{kj}(q_{kj}) - R_{ij}^z(q_{ij})F^0_{kj}(q_{kj}))
 \Big)+
     }
 \\ \ \\
  \displaystyle{
 +\delta_{ij}\Big( \nu^2{\sum\limits_{b,c}}'[F^0_{bc}(q_{bc}),r_{ic}(q_{ic})]+\nu{\sum\limits_{b,c}}'p_b F^0_{bc}(q_{bc})
 -\frac{\nu^2}{3}{\sum\limits_{a,b,c}}'[F^0_{ab}(q_{ab}),r_{cb}(q_{cb})]\Big)\,,
  }
 \end{array}
 \eq
 Two upper lines of (\ref{q32}) are straightforward generalizations
 of (\ref{q08}), while the last line is non-trivial for $\ti N>1$
 only (more precisely, for $\ti N=1$ it is proportional to the identity matrix).
  Its role is similar to the $\mF^0$ term in (\ref{q21}).
As in the case of the second flow here the Lax equations ${\dot
\mL}=[\mL,\mM^{(3)}]$ is equivalent to equations of motion
(\ref{q04}), where the coupling constant $\nu$ is replaced by $\ti
N\nu$. The proof is direct and somewhat technical. It uses
(\ref{q29}), (\ref{q30}) together with the classical Yang-Baxter
equation
 \beq\label{q33}
 \begin{array}{c}
  \displaystyle{
 [r_{ij}(q_{ij}),r_{ik}(q_{ik})]+[r_{ij}(q_{ij}),r_{jk}(q_{jk})]+[r_{ik}(q_{ik}),r_{jk}(q_{jk})]=0\quad
 \forall i,j,k
 }
 \end{array}
 \eq
or, to be exact, with its derivative
 \beq\label{q34}
 \begin{array}{c}
  \displaystyle{
 [F_{ij}^0(q_{ij}),r_{ki}(q_{ki})+r_{kj}(q_{kj})]=[F_{ik}^0(q_{ik}),r_{jk}(q_{jk})+r_{ji}(q_{ji})]
 =[F^0_{jk}(q_{jk}),r_{ij}(q_{ij})+r_{ik}(q_{ik})]\,.
 }
 \end{array}
 \eq
Details of the proof will be given elsewhere.


\paragraph{R-matrix-valued Lax pairs and spin chains.}
 \setcounter{section}{4}
 \setcounter{equation}{0}
We are now in a position to describe relationship between
$R$-matrix-valued Lax pairs and long-range spin chains. For this
purpose we restrict ourself to the equilibrium position (\ref{q14}).
Then the Lax matrix (\ref{q20}) turns into\footnote{We may put
$\nu=1$ since it is a common factor in the Lax equations.}
  \beq\label{q35}
  \begin{array}{c}
    \displaystyle{
{\mL}^{\rm chain}(z)=\sum\limits_{i,j=1}^N E_{ij}\otimes
(1-\delta_{ij})R^{\,z}_{ij}(x_{ij})\,.
 }
 \end{array}
 \eq
The restriction of the $M$-matrix (\ref{q21}) is subdivided into two
parts as $\mM^{(2)}=(\mM^{(2)}-\mF^0)+\mF^0$. The restriction of the
first term is
  \beq\label{q36}
  \begin{array}{c}
  \displaystyle{
\mM^{\rm chain}(z)=\sum\limits_{i,j=1}^N E_{ij}\otimes\Big(
-\delta_{ij}\sum\limits_{k\neq i}^N F^{\,0}_{ik}(x_{ik})
+(1-\delta_{ij})F^{\,z}_{ij}(x_{ij})\Big)\,,
 }
 \end{array}
 \eq
while the restriction of the second term is denoted as
  \beq\label{q37}
  \begin{array}{c}
  \displaystyle{
\mH_2^{\rm chain}
 =\sum\limits_{k>m}^N F^{\,0}_{km}(x_{km})\,.
 }
 \end{array}
 \eq
Then the restriction of the (classical) Lax equation for
Calogero-Moser model gives the quantum Lax equation for the spin
chain with the Hamiltonian $\mH_2^{\rm chain}$:
  \beq\label{q38}
  \begin{array}{c}
  \displaystyle{
[\mH_2^{\rm chain},{\mL}^{\rm chain}(z)]=[{\mL}^{\rm
chain}(z),{\mM}^{\rm chain}(z)]\,.
 }
 \end{array}
 \eq
This equation holds for an arbitrary $R$-matrix (entering $\mL$ and
$\mM^{(2)}$) which satisfies associative Yang-Baxter equation
(\ref{q23}) together with the unitarity and skew-symmetry properties
(\ref{q26}), (\ref{q261}).

The Inozemtsev chain (\ref{q12}), (\ref{q15})-(\ref{q17}) is
reproduced from (\ref{q35})-(\ref{q38}) as follows. Consider the
following $R$-matrix:
  \beq\label{q39}
  \begin{array}{c}
  \displaystyle{
R_{ij}^z(x_{ij})=P_{ij}\,\phi(z,x_{ij})
 }
 \end{array}
 \eq
It satisfies all necessary conditions. In particular, the
associative Yang-Baxter equation (\ref{q23}) for this $R$-matrix
follows from $P_{ab}P_{bc}=P_{ac}P_{ab}=P_{bc}P_{ac}$ and the scalar
Fay identity (\ref{q11}). Therefore, we may substitute it into
(\ref{q35})-(\ref{q37}). The corresponding analogues of the
classical $r$-matrix and its derivative are given by
  \beq\label{q40}
  \begin{array}{c}
  \displaystyle{
r_{ij}(x_{ij})=P_{ij}\,E_1(x_{ij})\,,\quad\quad
F^0_{ij}(x_{ij})=P_{ij}\,f(0,x_{ij})\,.
 }
 \end{array}
 \eq
As a result we obtain for the case (\ref{q39}): $\mL^{\rm
chain}(z)=\mL^{\rm Inoz}(z)$, $\mM^{\rm chain}(z)=-\mM^{\rm
Inoz}(z)$ and
  \beq\label{q41}
  \begin{array}{c}
  \displaystyle{
\mH_2^{\rm chain}\stackrel{(\ref{q071})}{=}
-\sum\limits_{i<j}P_{ij}\wp(x_{ij})+\frac{1}{3}\frac{\vth'''(0)}{\vth'(0)}\sum\limits_{i<j}P_{ij}=-H_2^{\rm
Inoz}+\frac{1}{3}\frac{\vth'''(0)}{\vth'(0)}\sum\limits_{i<j}P_{ij}\,.
 }
 \end{array}
 \eq
Notice that the spin chain Hamiltonian (\ref{q37}) appeared as a
part ($\mF^0$-term) of the $\mM^{(2)}$-matrix restricted to
$q_j=x_j$. The $\mF^0$-term enters $\mM^{(2)}$ as
$\nu1_N\otimes\mF^0$, hence it is (up to a number factor) equal to
trace of $\mM^{(2)}$-matrix over the auxiliary space, which is the
first (${\rm Mat}_N$-valued) tensor component in its definition:
  \beq\label{q42}
  \begin{array}{c}
  \displaystyle{
 \mH_2^{\rm chain}\propto\tr_{\rm
 aux}\,\mM^{(2)}\left.\right|_{ q_j=x_j}\propto\mF^0\left.\right|_{ q_j=x_j}
 }
 \end{array}
 \eq
The above mentioned arguments can be applied to the higher flows of
the Calogero-Moser model as well. For example, for the third flow
 $\mM^{(3)}$ (\ref{q32}) we have
  \beq\label{q43}
  \begin{array}{c}
  \displaystyle{
\tr_{\rm aux}\,\mM^{(3)}= N\nu{\sum\limits_{b,c}}'p_b
F^0_{bc}(q_{bc})
 +\nu^2\Big(1-\frac{N}{3}\Big){\sum\limits_{a,b,c}}'[F^0_{ab}(q_{ab}),r_{cb}(q_{cb})]
  }
 \end{array}
 \eq
 Recall that the prime means summation over all pairwise distinct values of indices.
After imposing constraints $p_i=0$, $q_i=x_i$ we are left with
  \beq\label{q441}
  \begin{array}{c}
  \displaystyle{
 \mH_3^{\rm chain}={\sum\limits_{a,b,c}}'[F^0_{ab}(x_{ab}),r_{cb}(x_{cb})]\,.
  }
 \end{array}
 \eq
 Then let us define the third Hamiltonian as
  \beq\label{q44}
  \begin{array}{c}
  \displaystyle{
 \mH_3^{\rm chain}={\sum\limits_{i<j<k}}[F^0_{ij}(x_{ij}),r_{ik}(x_{ik})+r_{jk}(x_{jk})]\,.
  }
 \end{array}
 \eq

We conjecture that the obtained in this way spin chain Hamiltonians
commute for some non-trivial (anisotropic) $R$-matrices. First,
consider the Inozemtsev case (\ref{q40}). A comparison of the poles and residues 
shows that in this case
  \beq\label{q45}
  \begin{array}{c}
  \displaystyle{
 \mH_3^{\rm
 chain}=-\frac{1}{36}\Big(J_2-\frac{1}{3}\frac{\vth'''(0)}{\vth'(0)}J_1\Big)\,,
  }
 \end{array}
 \eq
where $J_1$, $J_2$ are given by (\ref{q18}), (\ref{q19}).

For the following $R$-matrices the commutativity $[\mH_2^{\rm
chain},\mH_3^{\rm  chain}]=0$ can be verified numerically:

\noindent 1. Baxter's elliptic XYZ $R$-matrix ($\tau$ is the
elliptic moduli)
  \beq\label{q46}
  \begin{array}{c}
  \displaystyle{
 R^z_{12}(q) =
 }
 \\ \ \\
  \displaystyle{
=1\otimes 1\,\phi(q,z)+\sigma_1\otimes\sigma_1\,e^{\pi i
q}\phi(q,z+\frac{\tau}{2})
 +\sigma_2\otimes\sigma_2\,e^{\pi i q}\phi(q,z+\frac{\tau+1}{2})+
 \sigma_3\otimes\sigma_3\,\phi(q,z+\frac{1}{2})\,.
  }
 \end{array}
 \eq
Then the classical $r$-matrix
  \beq\label{q47}
  \begin{array}{c}
  \displaystyle{
 r_{12}(q)
=1\otimes 1\,E_1(q)+\sigma_1\otimes\sigma_1\,e^{\pi i
q}\phi(q,\frac{\tau}{2})
 +\sigma_2\otimes\sigma_2\,e^{\pi i q}\phi(q,\frac{\tau+1}{2})+
 \sigma_3\otimes\sigma_3\,\phi(q,\frac{1}{2})\,.
  }
 \end{array}
 \eq
For the three functions $\vf_1(q)=e^{\pi i
q}\phi(q,\frac{\tau}{2})$, $\vf_2(q)=e^{\pi i
q}\phi(q,\frac{\tau+1}{2})$ and $\vf_3(q)=\phi(q,\frac{1}{2})$ the
derivative of a one is given by the minus product of two others:
$\p_q\vf_\al(q)=-\vf_\be(q)\vf_\ga(q)$. Therefore, the second
Hamiltonian (\ref{q37}) acquires the form (\ref{qq024}):
  \beq\label{q48}
  \begin{array}{c}
  \displaystyle{
 \mH_2^{\rm  chain}
 =\sum\limits_{i<j}\Big(\stackrel{i}{\sigma}_0\stackrel{j}{\sigma}_0E_1'(x_{ij})-
 \sum\limits_{\al=1}^3\stackrel{i}{\sigma}_\al\stackrel{j}{\sigma}_\al\vf_\be(x_{ij})\vf_\ga(x_{ij})\Big)=
 }
 \\
  \displaystyle{
  =\sum\limits_{i<j}\Big(\stackrel{i}{\sigma}_0\stackrel{j}{\sigma}_0E_1'(x_{ij})+
 \sum\limits_{\al=1}^3\stackrel{i}{\sigma}_\al\stackrel{j}{\sigma}_\al\vf_\al(x_{ij})(E_1(x_{ij}+\om_\al)-E_1(x_{ij})-E_1(\om_\al))\Big)\,,
  }
 \end{array}
 \eq
where $\om_\al$ is the half-period (the second argument of
$\vf_\al(q)$). One more useful form for $\mH_2^{\rm  chain}$ is as
follows:
  \beq\label{q49}
  \begin{array}{c}
  \displaystyle{
 \mH_2^{\rm  chain}
 =\frac{N(N-1)}{6}\frac{\vth'''(0)}{\vth'(0)}\,\sigma_0^{\otimes N}
 -\frac12\sum\limits_{i<j}\Big(\sum\limits_{\al=0}^3\stackrel{i}{\sigma}_\al\stackrel{j}{\sigma}_\al\wp(\frac{x_{ij}}{2}+\om_\al)\Big)P_{ij}\,,
  }
 \end{array}
 \eq
where $\om_0=0$. The third Hamiltonian is evaluated through
(\ref{q44}).

\noindent 1. Trigonometric XXZ 6-vertex $R$-matrix
  \beq\label{q50}
  \begin{array}{c}
  \displaystyle{
 R_{12}^z(q) =(\pi \cot{\pi z} + \pi \cot{\pi q}) \cdot (\sigma_0 \ox \sigma_0 + \sigma_3 \ox \sigma_3) +
 }
 \\ \ \\
  \displaystyle{
 +\frac{\pi}{\sin{\pi z}} \cdot (\sigma_0 \ox \sigma_0 - \sigma_3 \ox \sigma_3)
+        \frac{\pi}{\sin{\pi q}} \cdot (\sigma_1 \ox \sigma_1 +
\sigma_2 \ox \sigma_2)\,.
  }
 \end{array}
 \eq
Then
  \beq\label{q51}
  \begin{array}{c}
  \displaystyle{
r_{12}(q) = \pi \cot{\pi q} \cdot (1 \ox 1 + \sigma_3 \ox \sigma_3)
+
        \frac{\pi}{\sin{\pi q}} (\sigma_1 \ox \sigma_1 + \sigma_2 \ox \sigma_2)
 }
 \end{array}
 \eq
and
  \beq\label{q52}
  \begin{array}{c}
  \displaystyle{
F^0_{12}(q) = -\frac{\pi^2}{\sin^2{\pi q}} \cdot (1 \ox 1 + \sigma_3
\ox \sigma_3 +
        \cos{\pi q} (\sigma_1 \ox \sigma_1 + \sigma_2 \ox
        \sigma_2))\,.
 }
 \end{array}
 \eq
This gives
  \beq\label{q53}
  \begin{array}{c}
  \displaystyle{
 \mH_2^{\rm  chain}=-\pi^2 \sum\limits_{i<j}
        \frac{\cos({\pi x_{ij}}) (\stackrel{i}{\sigma}_1 \stackrel{j}{\sigma}_1 + \stackrel{i}{\sigma}_2 \stackrel{j}{\sigma}_2) +
            \stackrel{i}{\sigma}_3 \stackrel{j}{\sigma}_3}{\sin^2(\pi x_{ij})} +
            C_N\sigma_0^{\otimes N}\,,\quad C_N=-\pi^2
            \sum\limits_{i<j}\frac{1}{\sin^2(\pi x_{ij})}\,.
  }
 \end{array}
 \eq
The third Hamiltonian (\ref{q44}) has compact form in this case:
  \beq\label{q54}
  \begin{array}{c}
  \displaystyle{
 \mH_3^{\rm  chain}=
 }
 \\ \ \\
  \displaystyle{
 -\frac{\pi^3}{2} {\sum\limits_{i<j<k}}\!
  \frac{\cos{\pi x_{ij}} (\stackrel{i}{\sigma}_1 \stackrel{j}{\sigma}_2 - \stackrel{j}{\sigma}_1 \stackrel{i}{\sigma}_2)\! \stackrel{k}{\sigma}_3 +
  \cos{\pi x_{jk}} (\stackrel{j}{\sigma}_1 \stackrel{k}{\sigma}_2 - \stackrel{k}{\sigma}_1 \stackrel{j}{\sigma}_2)\! \stackrel{i}{\sigma}_3 +
  \cos{\pi x_{ki}} (\stackrel{k}{\sigma}_1 \stackrel{i}{\sigma}_2 - \stackrel{i}{\sigma}_1 \stackrel{k}{\sigma}_2)\! \stackrel{j}{\sigma}_3}
              {\sin{\pi x_{ij}} \sin{\pi x_{jk}} \sin{\pi
              x_{ki}}}\,.
  }
 \end{array}
 \eq
 The spin exchange operator entering
(\ref{q53}) was obtained in \cite{HW2}
 in their study of the spin Calogero-Moser models, and the spin chains of
 this type were considered in \cite{Fukui,Beisert}.


\paragraph{Conclusion.}
 \setcounter{section}{4}
 \setcounter{equation}{0}

The purpose of the paper is two-fold. First, we study
$R$-matrix-valued Lax pairs for the classical Calogero-Moser model
and describe its third flow. Then we mention that the scalar part
(in the auxiliary space) of the $M$-matrices provides spin exchange
operators entering the Hamiltonians of the long range spin chains.
We conjecture commutativity $[\mH_2^{\rm chain},\mH_3^{\rm
chain}]=0$ for this Hamiltonians restricted to the equilibrium
position $p_i=0$, $q_i=x_i=i/N$. Such hypothesis is based on the
coincidence of these Hamiltonians with those for Inozemtsev chain
for a special choice of the $R$-matrix. For the Baxter's elliptic
$R$-matrix we verify numerically that these Hamiltonians commute. In
this way the anisotropic extension of the Inozemtsev chain is
described.

Let us also mention that the conjecture on commutativity
$[\mH_2^{\rm chain},\mH_3^{\rm  chain}]=0$ does not hold true for
any $R$-matrix satisfying associative Yang-Baxter equation (and
other properties). For instance, it is not true for the 7-vertex
$R$-matrix presented in \cite{Chered}. Another remark is that we
study $R$-matrices depending on the spectral parameter only. To
include the rest of $R$-matrices into the construction of
$R$-matrix-valued Lax pairs is a challenging task, since the
Haldane-Shastry type chains are known to exist for the quantum group
like $R$-matrices \cite{Uglov}.



\vskip2mm

\noindent {\bf Acknowledgments.} We are grateful to A. Grekov and N.
Slavnov for helpful discussions. The work was performed at the
Steklov Mathematical Institute of Russian Academy of Sciences,
Moscow. This work is supported by the Russian Science Foundation
under grant 14-50-00005.

\begin{small}

\end{small}

\end{document}